\documentclass[12pt]{article}
\pdfoutput=1
\usepackage[a4paper]{geometry}
\usepackage{amsmath,amssymb,amsfonts,latexsym,yhmath}
\usepackage{graphicx}  
\usepackage[colorlinks=true, pdfstartview=FitV, linkcolor=blue, citecolor=blue, urlcolor=blue]{hyperref}

\newcommand{\ep}{\qquad {\vrule height 10pt width 8pt depth 0pt}}

\newcommand {\bN}{{\mathbb N}}

\newcommand {\bZ}{{\mathbb Z}}
\newcommand {\bI}{{\mathbb I}}

\newcommand {\bC}{{\mathbb C}}

\newcommand {\bE}{{\mathbb E}}

\newcommand {\bH}{{\mathbb H}}
\newcommand {\bT}{{\mathbb T}}
\newcommand {\bP}{{\mathbb P}}

\newcommand {\cO}{{\cal O}}

\newcommand {\ii}{{\imath}}

\newtheorem{theorem}{Theorem} [section]

\newtheorem{propo}[theorem]{Proposition}

\newtheorem {remark}[theorem]{Remark}

\begin{document}
\title{Dynamical Localization of the Chalker-Coddington Model far from Transition}
\author{Joachim Asch
\thanks{
CPT, CNRS UMR 7332, Aix--Marseille Universit\'e et 
Universit\'e du Sud, ToulonÐVar, BP 20132,
F--83957 La Garde Cedex, France, e-mail:
asch@cpt.univ-mrs.fr},
Olivier Bourget
\thanks{
Departamento de Matem\'aticas
Pontificia Universidad Cat\'olica de Chile, Av. Vicu\~{n}a Mackenna 4860,
C.P. 690 44 11, Macul
Santiago, Chile},
Alain Joye
\thanks{
UJF-Grenoble, CNRS Institut Fourier UMR 5582, Grenoble, 38402, France}
}
\date{16.12.2011}
\maketitle

\abstract{We study a quantum network percolation model which is numerically pertinent to the understanding of the delocalization transition of the quantum Hall effect. We show dynamical localization for parameters corresponding to edges of Landau bands, away from the expected transition point.}

\section{Introduction, the model and the results}\label{sec:intro}

The Chalker - Coddingtion effective model was introduced  in \cite{cc} in order to study the quantum Hall transition numerically in a quantitative way, see \cite{kok} for a review.  Main features of the dynamics of a 2D electron in a strong perpendicular magnetic field and a smooth bounded random potential are described by a  random unitary  ${U}$ acting on $l^{2}(\bZ^{2})$.

In this effective picture the $\bZ^{2}$ lattice points label the directed edges of a graph on which the electron moves. These sites communicate with some of their nearest neighbors by superpositions of tunneling to different directions with real amplitudes $r$ and $t$, such that $r^{2}+t^{2}=1$ and random phases. There is no backscattering. The folklore intuition is that for $\vert r\vert\neq\vert t \vert$ the localization length is finite whereas for $\vert r\vert=\vert t\vert$ the system delocalizes; this defines the transition point. The parameter 
$\vert t\vert$ is $\frac{1}{\sqrt{1+e^{\varepsilon}}}$ where $\varepsilon$ is the distance of the electrons energy  to the nearest Landau Level. An application of  a numerical finite size scaling method  led   Chalker and Coddington \cite{cc}, see also \cite{kok}, to conjecture that the localization length diverges as $\vert t/r\vert \to1$ as 
\[\left(\frac{1}{\ln\vert\frac{t}{r}\vert}\right)^{\alpha}\]
where the critical exponent  $\alpha$ exceeds substantially the exponent expected when a classical percolation model is applied to the magnetic random propagation problem. The values advocated for $\alpha$ are $2.5\pm0.5$ for the quantum and $4/3$ for the classical case, \cite{t,kok}. 
See \cite{abj,am} for more information and literature on this model. 
While the interest of the Chalker Coddington model lies in the transition point $\vert t\vert=\vert r\vert$ remark that only very few results on delocalization are known, see \cite{k}, \cite{aw}.  

We present here results on localization for situations which are near the integrable cases $rt=0$. A  large amount of precise information on localization in random media is available. In the selfadjoint case, these are obtained by the multiscale method \`a la \cite{fs}, see also \cite{s}, or the fractional moment method \`a la \cite{aim}, see also \cite{aenss} from which derive the results in the unitary case, see \cite{hjs}, which we use here. 

To define the model consider the angle $\varphi$ given by $\left(\cos\varphi,\sin\varphi\right):=\left(t,r\right)$,  the family of random unitaries is
\begin{equation}U_{\omega}(\varphi)=D_{\omega}S(\varphi)\qquad \hbox{ \rm on } l^{2}(\bZ^{2})\label{def:u}\end{equation}
where the matrix of the unitary $D_{\omega}$ is diagonal in the standard basis with entries uniformly and independently distributed on  $\bT$, the complex circle of radius one.
More precisely:  for $\omega\in\bT^{\bZ^{2}}$ considered as a probability space with $\sigma$ algebra generated by the cylinder sets and measure $\bP=\otimes_{\mu\in\bZ^{2}}d\ell$ where $d\ell$ is the normalized Lebesque measure on $\bT$: $\left(D_{\omega}\right)_{\mu\nu}=\omega_{\mu}\delta_{\mu\nu}$.

$S$ is the deterministic unitary 
\[S(\varphi):=\cos\varphi S_{\circlearrowleft}+i \sin\varphi S_{\circlearrowright} \] 
built by superposition of local (anti-)clockwise rotations in the following sense: for the the standard basis $\{e_{\mu}\}_{\mu\in\bZ^{2}}$ of $l^{2}\left(\bZ^{2}\right)$, $\left(e_{\mu}\right)_{\nu}=\delta_{\mu\nu}$, consider the decompositions
\[\bigoplus_{j,k\in\bZ}\bH_{\circlearrowleft}^{j,k}=l^{2}\left(\bZ^{2}\right)=\bigoplus_{j,k\in\bZ}\bH_{\circlearrowright}^{j,k}\]
where
\[\bH_{\circlearrowleft}^{j,k}:=span\left\{e_{(2j,2k)}, e_{(2j+1,2k)},e_{(2j+1,2k+1)}, e_{(2j,2k+1)} \right\},\]
\[\bH_{\circlearrowright}^{j,k}:=span\left\{e_{(2j,2k)}, e_{(2j,2k-1)},e_{(2j-1,2k-1)}, e_{(2j-1,2k)} \right\}.\]

Then
\[S_{\circlearrowleft}:=\bigoplus_{j,k\in\bZ} S_{\circlearrowleft}^{j,k}, \qquad S_{\circlearrowright}:=\bigoplus_{j,k\in\bZ} S_{\circlearrowright}^{j,k}\]

where for $\#\in\lbrace\circlearrowleft,\circlearrowright\rbrace$ the restrictions $S_{\#}^{j,k}$ of $S_{\#}$ to the invariant subspaces $\bH_{\#}^{j,k}$ are represented with respect to their basisvectors in the above indicated order by the permutation matrix
\[
\left(\begin{array}{cccc}
 0 & 0  & 0 & 1  \\
 1 & 0  & 0 & 0  \\
  0 & 1  & 0 & 0 \\
   0 & 0  & 1 & 0    \end{array}\right),\]
 
 i.e. $S_{\circlearrowleft}^{j,k}e_{2j,2k}=e_{2j+1,2k}\ldots$.
 
 \begin{figure}[t]
\centerline{
\includegraphics[width=4cm]{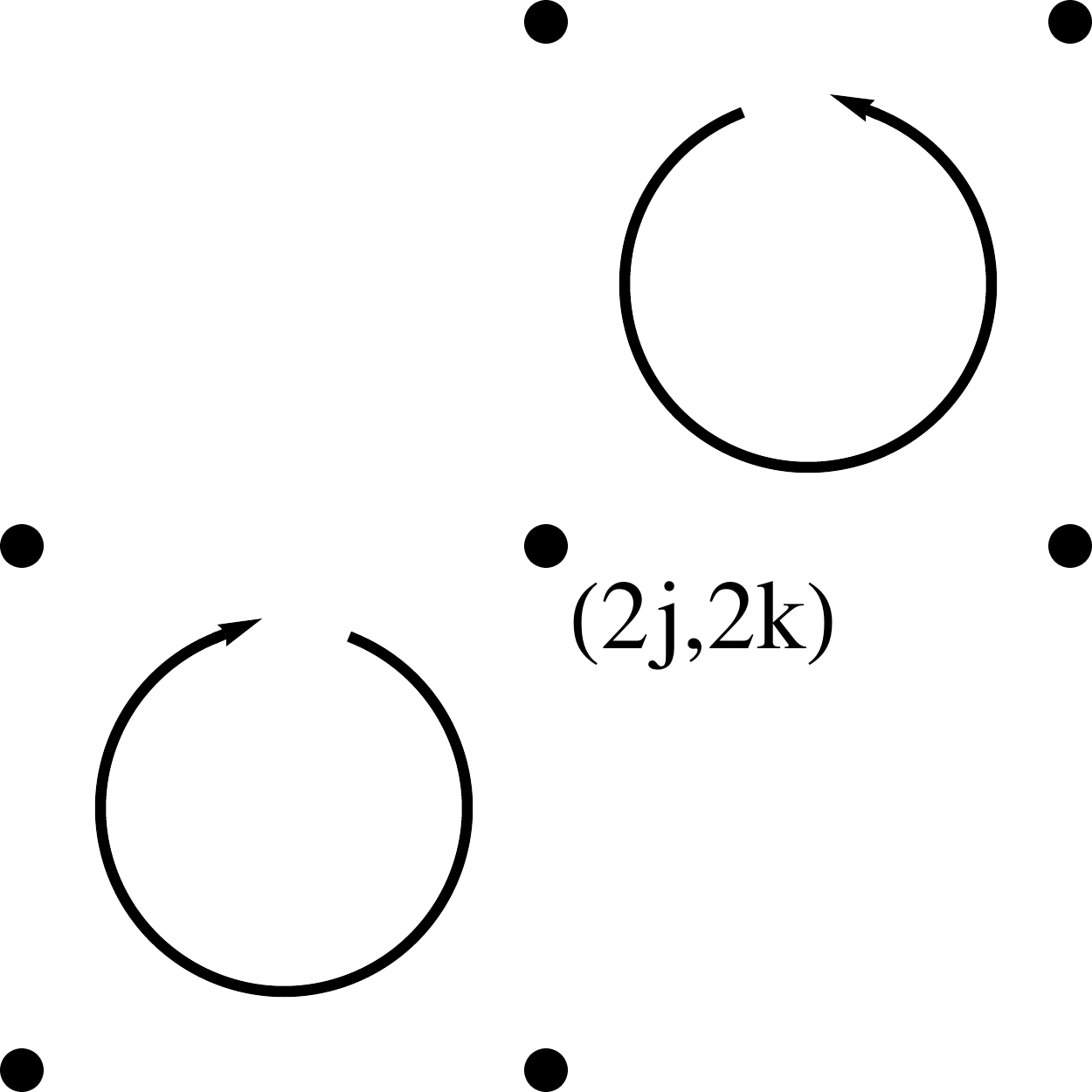}
}
 \caption{The action of $S_{\circlearrowleft}^{j,k}$ and $S_{\circlearrowright}^{j,k}$}
 \end{figure}

Remark that due to the distribution of the phases and the invariance of $S(\varphi)$ with respect to the action of $\Theta$ of $\bZ^{2}$ on $l^{2}(\bZ^{2})$ defined by $\Theta_{\nu}(\psi)(\mu):=\psi(\mu+2\nu)$ we have
\[\Theta_{\nu}U_{\omega}(\varphi)\Theta_{\nu}^{-1}=U_{\Theta_{\nu}\omega}(\varphi)\]
thus  for each $\varphi$,  $U_{\omega}(\varphi)$ is an ergodic family of random unitaries. Our main result is

\begin{theorem}{\label{thm:main}}
There exists a $\varphi_{0}>0$ such that for $\vert\varphi\mod \frac{\pi}{2}\vert\le\varphi_{0}$ it holds for the familiy $U_{\omega}(\varphi)$ defined in (\ref{def:u}) :
\begin{enumerate}
\item $U_{\omega}(\varphi)$ has pure point spectrum almost surely;
\item for any nonnegative $p$ and any $\psi\in l^{2}(\bZ^{2})$ of compact support it holds for the multiplication operator $\vert X\vert^{p}e_{\mu}=\vert\mu\vert^{p}e_{\mu}:=\left(\sum_{i=1}^{2}\mu_{i}^{2}\right)^{p/2}e_{\mu}$ almost surely:
\[\sup_{n\in\bZ}\left\Vert \vert X\vert^{p}U^{n}_{\omega}(\varphi)\psi\right\Vert<\infty;\]
\item there exist positive constants $g, c$ such that for all $\mu, \nu\in\bZ^{2}$
\[\bE\left\lbrack\sup_{f\in C(\bT), \Vert f\Vert_{\infty}\le1}\left\vert\langle e_{\mu},f\left(U_{\omega}(\varphi)\right) e_{\nu}\rangle\right\vert\right\rbrack\le c e^{-g\vert\mu-\nu\vert}\]
where $C(\bT)$ are the complex valued continuous functions on the circle.
\end{enumerate}
\end{theorem}

\begin{remark} The strongest result is dynamical localization (3) which implies  exponential localization (2) and spectral localization (1).\end{remark}

In the proof we will consider finite subspaces of $l^{2}(\bZ^{2})$ built of finite sums of blocks $\bH^{j,k}_{\circlearrowleft}$ and unitary restrictions of $U$ imposing elastic boundary conditions. A corollary of the method is that the above results hold also for unitary restrictions to strips of finite but arbitrary width, see theorem \ref{thm:strip} stated below. 


In contrast remark that in a previous paper \cite{abj} we proved  spectral localization for the restriction of the model to a strip of width $2M$ and periodic boundary conditions as well as a Thouless formula for any $\varphi$.

Our strategy to prove theorems \ref{thm:main} and \ref{thm:strip} is to show that the ``localization machinery'' as exposed in \cite{hjs} for the case of random unitaries applies to the Chalker Coddington model. To do so we have to consider the restriction of $U_{\omega}(\varphi)$ to boxes of finite size and to control the probability of occurrence of spectral gaps on a suitable scale. This is done in section (\ref{sec:eigenvalues}). In section (\ref{sec:proof}) we then finish the proof implementing an iteration procedure based on a resampling argument. Remark that a simpler strategy based on the use of a decoupling lemma \`a la \cite{aim}, see \cite{j} for the unitary case, cannot be used here because the deterministic part $S$ is purely is purely off--diagonal.

\section{Resolvent estimates for restriction to finite regions }\label{sec:eigenvalues}
Whereas theorem \ref{thm:main} is stated for the cases either $\cos\varphi$ or $\sin\varphi$ small enough, resp. theorem \ref{thm:strip} for the case $\sin\varphi$ small,  we will explicit the proof for $\vert\varphi\vert$ small enough. The other cases can be treated analogously. The restriction to finite regions is conditioned by this.

\medskip
We now define unitary restrictions of $U$ to regions of finite volume. Then we control the probability of small spectral gaps for growing volume. This is a major ingredient of the proof of localization.

\begin{figure}[hb]
\centerline{
\includegraphics[width=.85\textwidth]{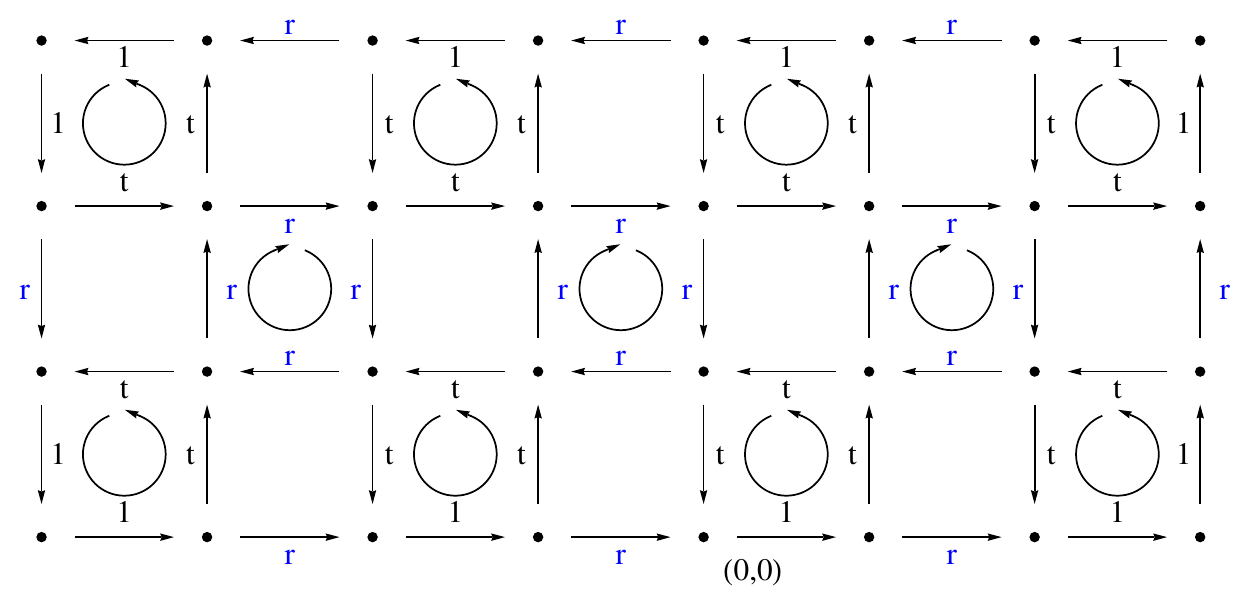}
}
 \caption{The action of $S^{\Lambda_{L}}$ for $L=(2,1)$}
 \label{fig:restriction}
 \end{figure}

For $L:=(L_{1},L_{2})\in\bN^{2}$ and
\[\Lambda_{L}:=\bZ^{2}\cap\left(\lbrack-2L_{1},2L_{1}-1\rbrack\times\lbrack-2L_{2}+2,2L_{2}+1\rbrack\right)\]
define ${\rm vol}\Lambda_{L}:=4L_{1}L_{2}$ and a unitary restriction  $U^{\Lambda_{L}}$ to the sum of  $4L_{1}L_{2}$ blocks. 
\[U_{\omega}^{\Lambda_{L}}(\varphi)\quad\hbox{ \rm on }\quad L^{2}\left(\Lambda_{L}\right):=\bigoplus_{\substack{ j\in\lbrack-L_{1},L_{1}-1\rbrack\\
k\in\lbrack-L_{2}+1,L_{2}\rbrack}}
\bH_{\circlearrowleft}^{j,k}\]
\[U_{\omega}^{\Lambda_{L}}(\varphi):=D_{\omega}^{\Lambda_{L}}S^{\Lambda_{L}}(\varphi)\]

where $D_{\omega}^{\Lambda_{L}}$ is the restriction of $D_{\omega}$ to $L^{2}(\Lambda_{L})$ and
\[S^{\Lambda_{L}}(\varphi):=\chi_{\Lambda_{L}}S(\varphi)\chi_{\Lambda_{L}}+T^{\Lambda_L}(\varphi)\]
where $\chi_{\Lambda_{L}}$ denotes the multiplication by the characteristic function of $\Lambda_{L}$ and
\begin{eqnarray*}&&T^{\Lambda_L}(\varphi):=(1-\cos(\varphi))\times\\
&\phantom{-}&\left(\sum_{j=-L_{1}}^{L_{1}-1}\left(| 2j,2L_{2}+1\rangle\langle2j+1,2L_{2 }+1|+|2j+1,-2L_{2}+2\rangle\langle2j,-2L_{2}+2|\right)\right.\\
&&+\left.\sum_{k=-L_{2}+1}^{L_{2}}\left(| 2L_{1}-1,2k+1\rangle\langle2L_{1}-1,2k|+|-2L_{1},2k\rangle\langle-2L_{1},2k+1|\right)\right)\end{eqnarray*}
meaning that clockwise components are reflected at and thus completely transmitted along the walls (i.e.: $t$ is replaced by one along the walls), c.f. figure (\ref{fig:restriction}). Note that
\begin{eqnarray}\Vert T^{\Lambda_L}(\varphi)\Vert&\le& \max\left(\sup_{\mu}\sum_{\nu}\vert\langle e_{\mu},T^{\Lambda_L}(\varphi)e_{\nu}\rangle\vert ,\sup_{\nu}\sum_{\mu}\vert\langle e_{\mu},T^{\Lambda_L}(\varphi)e_{\nu}\rangle\vert \right)\nonumber\\
&\le&2\vert 1-\cos\varphi\vert.\label{eq:restimate}
\end{eqnarray}

For $M\in\bN$ denote

\begin{equation}
\label{def:lambdaInfty}
\Lambda_{(\infty,M)}:=\bZ^{2}\cap\left(\bZ\times \lbrack-2M+2,2M+1\rbrack\right) \quad \hbox{\rm and}\quad
U_{\omega}^{\Lambda_{(\infty,M)}}(\varphi)
\end{equation}
the corresponding unitary restriction.

 For $\varphi=0$ the dynamics decouples to blockwise--anticlockwise rotation and the spectrum $\sigma(U_{\omega}(0))$ equals $\bT$ almost surely. Indeed, for the restriction to $\bH_{\circlearrowleft}^{j,k}$ it holds:
\[\left(U_{\omega}^{j,k}(0)\right)^{4}=\left(D_{\omega}^{j,k}S_{\circlearrowleft}^{j,k}\right)^{4}=\left(\det D_{\omega}^{j,k}\right)\bI_{\bH_{\circlearrowleft}^{j,k}},\]
the random variables $\det D_{\omega}^{j,k}$ are uniform and  i.i.d, see Lemma 4.1 of [ABJ], thus the eigenvalues of $U_{\omega}^{j,k}(0)$ are a uniform i.i.d. random variable times the forth roots of unity. This observation allows us to estimate the probability of spectral gaps for $U_{\omega}^{\Lambda_{L}}(0)$.

We can now formulate our result about the strip mentioned above.

\begin{theorem}{\label{thm:strip}}
For any $M\ge1$ and $\vert\varphi \mod \pi\vert$ small enough the results of the theorem \ref{thm:main} hold true for the unitary restriction $U_{\omega}^{\Lambda_{(\infty,M)}}(\varphi)$ as defined in (\ref{def:lambdaInfty}).
\end{theorem}

We will prove this at the same time as theorem \ref{thm:main}. Let us remark that the smallness restrictions are due to the method of proof and we expect the result to be true for any $\varphi$ as in the case of periodic boundary conditions treated in \cite{abj}. An analogous  theorem  holds also for the other cases $\vert\varphi\mod\frac{\pi}{2}\vert$ small enough. 

\bigskip

For $(2k_{1}, 2k_{2})\in(2\bZ)^{2}$ we use the notation $\Lambda_{L}+(2k_{1},2k_{2})$ for the shifted box.

\begin{propo}\label{thm:gaps}
For any $L\in\bN^{2},\eta>0$ it holds
\[\bP\left({\rm dist}\left(z,\sigma\left(U_{\omega}^{\Lambda_{L}+v}(0)\right)\right)\le\eta\right)=\cO\left(\eta {\rm vol}\Lambda_{L})\right)\]
for $\eta{\rm vol}\Lambda_{L}$ small enough, uniformly in $v\in(2\bZ)^{2}$ and  $z\in\bC\setminus\bT$.
\end{propo}

\noindent {\bf Proof:}
By ergodicity it is sufficient to prove the claim for $v=0$.
For an arc $A\subset\bT$ of measure $\ell(A)<\frac{1}{4}$ the probability for the spectrum of $U_{\omega}^{j,k}(0)$ to lie outside of $A$ equals  $\left(1-4\ell(A)\right)$. $\Lambda_{L}$ contains $4L_{1}L_{2}$ blocks  $\bH_{\circlearrowleft}^{j,k}$, thus
\[\bP\left(\sigma\left(U^{\Lambda_{L}}_{\omega}(0)\right)\cap A=\emptyset\right)=\left(1-4\ell(A)\right)^{4L_{1}L_{2}}.\]

The spectrum is a subset of $\bT$. By a trigonometric estimate the intersection of $\bT$ and a ball of radius $\eta<1$ around $z$,  $B_{\eta}(z)\cap\bT$, has measure less then $\frac{\eta}{2}$ thus:

\[\bP\left(\sigma\left(U^{\Lambda_{L}}_{\omega}(0)\right)\cap B_{\eta}(z)=\emptyset\right)\ge\left(1-2\eta\right)^{4L_{1}L_{2}}.\]

It follows that
\[\bP\left(
{\rm dist}\left(
z,\sigma\left(U^{\Lambda_{L}}_{\omega}(0)\right)
\right)>
\eta
\right)\ge
\left(1-2\eta\right)^{4L_{1}L_{2}}.\]
Now we use the estimate 
\[\sup_{y\in(0,1/2)}\left\vert (1- x y)^{\frac{1}{y}}-(1-x)\right\vert\le\frac{x^{2}}{2}\qquad\forall  x\in[0,1]\]
with $y=\frac{1}{4L_{1}L_{2}}$ and $x={4 L_{1}L_{2}}2\eta=2{\rm vol}\Lambda_{L}\eta$ to conclude.
\hfill\ep

\bigskip

 Denote resolvents by $R$, e.g.
 \[R_{\omega}^{\Lambda_{L}}(\varphi,z):=\left(U_{\omega}^{\Lambda_{L}}(\varphi)-z\right)^{-1}\]
 for $\quad z\in\rho\left(U_{\omega}^{\Lambda_{L}}(\varphi)\right)$, the resolvent set.
 
We now prove that expectations of the elements of the boxed resolvent matrix are polynomially small in $\frac{1}{\vert L\vert}$.
\begin{propo}\label{propo:resolventestimate}
For  $s\in(0,1)$, $a\ge0$ 
\[\bE\left(\left\vert\langle e_{\mu},R^{\Lambda_{L}+v}(\varphi,z) e_{\nu}\rangle\right\vert^{s}\right)=\cO\left(\frac{1}{\vert L\vert^{a}}\right)\]
for each $p$ such that $1-\frac{1}{p}>s$, uniformly in the region $L\in\bN^{2}, z\in\bC\setminus\bT$, \\$\vert\varphi\vert\le\frac{1}{\vert L\vert^{2(a p+2)+\frac{a}{s}}}, v\in\bZ^{2}$, $\mu,\nu\in\Lambda_{L}+v$, $\vert\mu-\nu\vert_{\infty}\ge2$.
 \end{propo}

{\bf Proof.} By ergodicity it is sufficient to prove the claim for $v=0$. For $a=0$ the estimate holds without smallness assumption on $\varphi$. Indeed by a  result based on spectral averaging proven as Theorem 3.1 in \cite{hjs} which holds for very general unitaries  of the form $D_{\omega}S$ with $D_{\omega}$ diagonal and $S$ deterministic, banded and shift invariant we have  : $\forall\  s\in(0,1)\ \forall\ \varphi\ \exists\ c>0\ \forall\ \mu, \nu\ \forall\ z\in\bC\setminus\bT$:
\begin{equation}\label{eq:momentestimate}
 \bE\left(\left\vert\langle e_{\mu},R^{\#}_{\omega}(\varphi,z) e_{\nu}\rangle\right\vert^{s}\right)\le c
\end{equation}
where $R^{\#}$ stands either for the full resolvent $R$ or for $R^{\Lambda_{L}}$, any $L$. For $a>0$ we use the invariance of the spaces $\bH^{j,k}$ by $U(0)$, first order perturbation theory and proposition (\ref{thm:gaps}).

$\alpha\sim\beta$ denotes that $\alpha, \beta\in\bZ^{2}$ are in the same invariant subspace :
\[\alpha\sim\beta:\Longleftrightarrow\exists\ \bH_{\circlearrowleft}^{j,k} \hbox{ \rm such that } e_{\alpha},e_{\beta}\in\bH_{\circlearrowleft}^{j,k};\]
remark that
\[\langle e_{\alpha},R_{\omega}^{\Lambda_{L}}(0,z)e_{\beta}\rangle=0 \hbox{ \rm if } \alpha\nsim\beta. \]
From the resolvent identity 
\[R^{\Lambda_{L}}(\varphi)=R^{\Lambda_{L}}(0)+R^{\Lambda_{L}}(\varphi)\left(U^{\Lambda_{L}}(0)-U^{\Lambda_{L}}(\varphi)\right)R^{\Lambda_{L}}(0),\]
the fact that non nearest neighbors are not coupled by $U(\varphi)$, i.e.:
\[\langle e_{\alpha},U^{\Lambda_{L}}_{\omega}(\varphi)e_{\beta}\rangle=0\quad \hbox{ \rm if } \vert\alpha-\beta\vert_{\infty}>1\]
and the estimate
\[\Vert U^{\Lambda_{L}}_{\omega}(\varphi)-U^{\Lambda_{L}}_{\omega}(0)\Vert\le3\vert \cos\varphi-1\vert+\vert\sin\varphi\vert\le4\vert\varphi\vert,\]
we see that for $\mu,\nu\in\Lambda_{L}$, $\vert\mu-\nu\vert_{\infty}\ge2$, $z\in\bC\setminus\bT$

\begin{equation}\label{eq:4}
\begin{split}
&\left\vert\langle e_{\mu},R^{\Lambda_{L}}_{\omega}(\varphi,z) e_{\nu}\rangle\right\vert\le \\
&\sum_{\substack{\alpha,\beta\in\bZ^{2}, \beta\sim\nu\\\vert\alpha-\beta\vert_{\infty}=1}}\left\vert\langle e_{\mu},R^{\Lambda_{L}}_{\omega}(\varphi,z) e_{\alpha}\rangle\langle e_{\alpha},\left(U_{\omega}(\varphi)-U_{\omega}(0)\right) e_{\beta}\rangle\langle e_{\beta},R^{\Lambda_{L}}_{\omega}(0,z) e_{\nu}\rangle\right\vert\\
&\le c\vert\varphi\vert\frac{1}{{\rm dist}\left(z,\sigma\left(U_{\omega}^{\Lambda_{L}}(0)\right)\right)}\sup_{\substack{\alpha,\beta\in\bZ^{2}, \beta\sim\nu\\\vert\alpha-\beta\vert_{\infty}=1}}\left\vert\langle e_{\mu},R^{\Lambda_{L}}_{\omega}(\varphi,z) e_{\alpha}\rangle\right\vert,
\end{split}
\end{equation}
where $c$ is a numerical constant as the number of sites in the above sum is finite, independent of $L$.

Denote for $p>1$ and $z\in\bC\setminus\bT$ the events for which $z$ lies in a gap of length bigger than ${2\eta}$
\[G_{\eta}(z):=\left\lbrace\omega\in\bT^{\bZ^{2}}, {\rm dist}\left(z,\sigma\left(U_{\omega}^{\Lambda_{L}}(0)\right)\right)>\eta\right\rbrace\]
and $G^{c}_{\eta}(z)$ its complement. Remark that by proposition (\ref{thm:gaps}) 
\[\bP\left(G^{c}_{\eta}(z)\right)=\cO\left(\eta{\rm vol}\Lambda_{L}\right).\]
Denote by $\chi_{A}$ the characteristic function of the set $A$.
Now for  $\frac{1}{p}+\frac{1}{q}=1$ we have by H\"older's inequality and estimate (\ref{eq:momentestimate}) with $\frac{1}{q}>s$ :
\begin{equation}
\begin{split}
\bE&\left(\chi_{G_{\eta}^{c}(z)}\left\vert\langle e_{\mu},R^{\Lambda_{L}}_{\omega}(\varphi,z) e_{\nu}\rangle\right\vert^{s}\right)\le\\
&\bP\left(G_{\eta}^{c}(z)\right)^{\frac{1}{p}}\bE\left(\left\vert\langle e_{\mu},R^{\Lambda_{L}}_{\omega}(\varphi,z) e_{\nu}\rangle\right\vert^{sq}\right)^{\frac{1}{q}}=\cO\left(\left(\eta {\rm vol}\Lambda_{L}\right)^{\frac{1}{p}}\right).
\end{split}
\end{equation}
We fix the scale to $\eta:=\frac{1}{{\rm vol}\Lambda_{L}\vert L\vert^{ap}}$ thus the claim is proved on the set $G_{\eta}^{c}(z)$ where $z$ is ``close'' to the spectrum. 

In the worst case we have ${\rm vol}\Lambda_{L}\le2\vert L\vert^{2}$.
By perturbation theory, it holds if $4 \vert\varphi\vert<\eta/2$: 
\[{\rm dist}\left(\sigma\left(U^{\Lambda_{L}}_{\omega}(0)\right),z\right)>\eta\Longrightarrow {\rm dist}\left(\sigma\left(U^{\Lambda_{L}}_{\omega}(\varphi)\right),z\right)>\frac{\eta}{2}\]

thus, by the inequality (\ref{eq:4})  for $\vert\varphi\vert\le\frac{1}{\vert L\vert^{2(a p+2)+\frac{a}{s}}}$ and our choice for $\eta$ we can estimate the complementary part

\begin{equation}
\begin{split}
\chi&_{G_{\eta}(z)}(\omega)\left\vert\langle e_{\mu},R^{\Lambda_{L}}_{\omega}(\varphi,z) e_{\nu}\rangle\right\vert^{s}\le\\
&\chi_{G_{\eta}(z)}(\omega)\left\vert\frac{c\varphi}{{\rm dist}\left(z,\sigma\left(U_{\omega}^{\Lambda_{L}}(0)\right)\right){\rm dist}\left(z,\sigma\left(U_{\omega}^{\Lambda_{L}}(\varphi)\right)\right)}\right\vert^{s}\\
&\phantom{skip}\le\frac{2 c}{\vert L\vert^{a}}
\end{split}
\end{equation}
and thus
\[\bE\left(\chi_{G_{\eta}(z)}\left\vert\langle e_{\mu},R^{\Lambda_{L}}_{\omega}(\varphi,z) e_{\nu}\rangle\right\vert^{s}\right)=\cO\left(\frac{1}{\vert L\vert^{a}}\right).
\]

which finishes the proof of the proposition.
\ep

 \section{The iteration procedure}\label{sec:proof}
 
In order to  prove that matrix elements decay exponentially in the distance between the states we shall use geometric resolvent estimates relating the resolvent of the full system to the one decoupled along the borders of $\Lambda_{L}$.

\begin{figure}[h]\label{fig:decoupled}
\centerline{
\includegraphics[width=.8\textwidth]{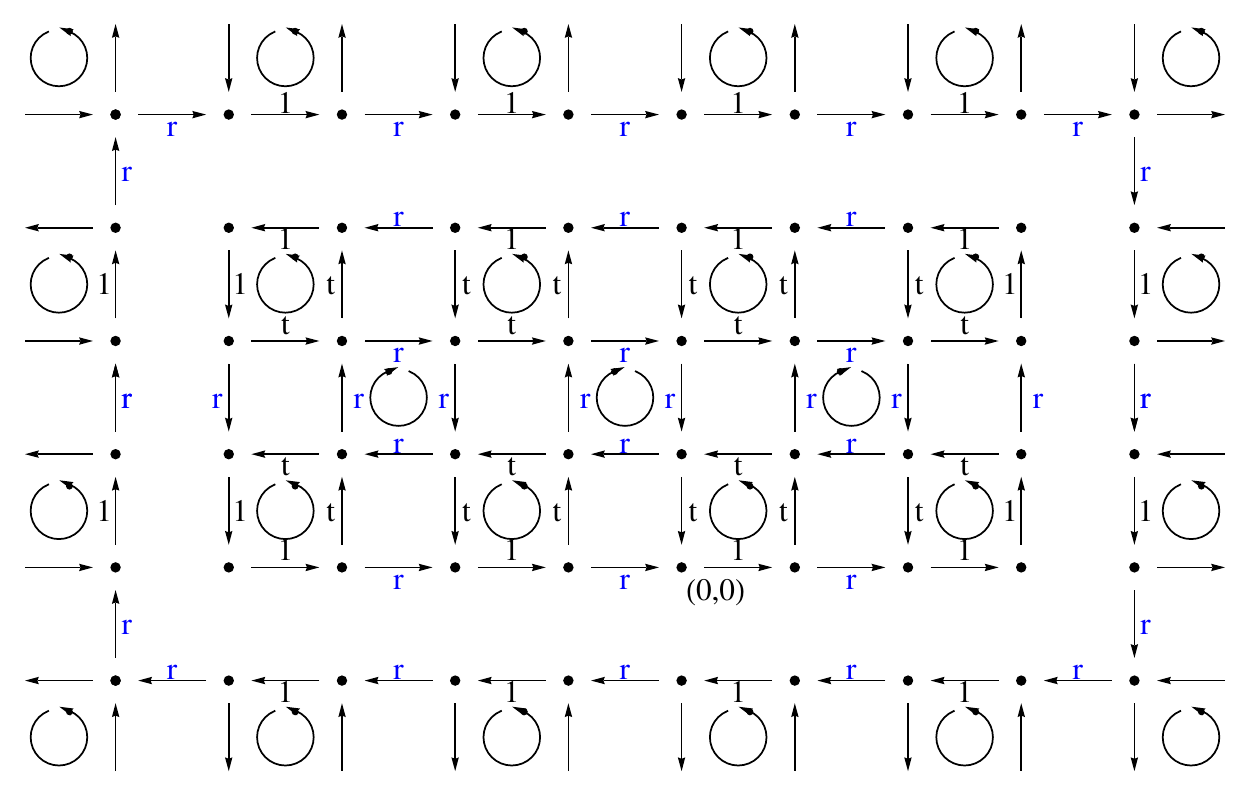}
}
 \caption{The action of $U^{(L)}$ for $L=(2,1)$}
 \end{figure}

Denote $\Lambda_{L}^{c}:=\bZ^{2}\setminus\Lambda_{L}$ and $S^{\Lambda_{L}^{c}}(\varphi)$ the unitary restriction of $S(\varphi)$ to $\Lambda_{L}^{c}$ such that the clockwise components are completely transmitted along the walls of $\Lambda_{L}^{c}$, c.f. the analogous definition of $S^{\Lambda_{L}}(\varphi)$. In the same spirit we construct $U^{\Lambda_{L}^{c}}_{\omega}(\varphi)=D^{\Lambda_{L}^{c}}_{\omega}S^{\Lambda_{L}^{c}}(\varphi)$. Then, see figure (\ref{fig:decoupled}),
\begin{equation}\label{def:ul}
U_{\omega}(\varphi):=\underbrace{U^{\Lambda_{L}}_{\omega}(\varphi)\oplus U^{\Lambda_{L}^{c}}_{\omega}(\varphi)}_{=:U_{\omega}^{(L)}(\varphi)}+V^{(L)}_{\omega}(\varphi)
\end{equation}

where each of the $\cO(\vert L\vert)$ non zero matrix elements of $V^{(L)}_{\omega}(\varphi)$ is $\cO(\vert\varphi\vert)$ uniformly in the parameters.
 
Define for $\mu\in\bZ^{2}$, $\lbrack\mu\rbrack$ the unique $\lbrack\mu\rbrack\in(2\bZ)^{2}$ (i.e.: the two components are even) such that $\lbrack\mu\rbrack\sim\mu$ and
\[\stackrel{\circ}{\Lambda}_{L}:=\left\lbrace\alpha\in\Lambda_{L}; \alpha+v\in\Lambda_{L} \forall v\in\bZ^{2}, \Vert v\Vert_{\infty}=1\right\rbrace,\qquad \partial\Lambda_{L}:=\Lambda_{L}\setminus\stackrel{\circ}{\Lambda}_{L},\]
and the nearest neighborhood of a subset of $\bZ^{2}$
\[N(S):=\left\lbrace\beta\in\bZ^{2};\beta=\alpha+v \hbox{ \rm for } \alpha\in S, v\in\bZ^{2}, \Vert v\Vert_{\infty}=1\right\rbrace.\]

For the full resolvent we prove:
\begin{propo}\label{propo:exponential}
For $s\in(0,1)$ there exist $\ell_{0}>0, \varphi_{0}>0$ and $q<1$ such that 
\[\bE\left(\left\vert\langle e_{\mu}, R(\varphi,z) e_{\nu}\rangle\right\vert^{s}\right)\le q \max_{\beta\in N\partial\left((\Lambda_{L}+\lbrack\mu\rbrack)^{c}\right)}\bE\left(\left\vert\langle e_{\beta}, R(\varphi,z) e_{\nu}\rangle\right\vert^{s}\right)\]
for all $L\in\bN^{2}, \vert L\vert>\ell_{0}$, $z\in\bC\setminus\bT, \vert\varphi\vert<\varphi_{0}, \mu\in\bZ^{2}, \nu\in\stackrel{\circ}{{\overbrace{(\Lambda_{L}+\lbrack\mu\rbrack)^{c}}}}$.\end{propo}

{\bf Proof.}  By ergodicity it is sufficient to prove the claim for $\mu=0$. Define $\ii:=(1,1)$.
For $V^{(L)}$ defined in (\ref{def:ul}) it holds
\[0=\chi_{\stackrel{\circ}{\Lambda}_{L}}V^{(L)}=V^{(L)}\chi_{\stackrel{\circ}{\Lambda}_{L}}=\chi_{\stackrel{\circ}{\Lambda_{L}^{c}}}V^{(L)}=V^{(L)}\chi_{\stackrel{\circ}{\Lambda_{L}^{c}}}.\]

\begin{figure}[h]\label{fig:iota}
\centerline{
\includegraphics[width=.32\textwidth]{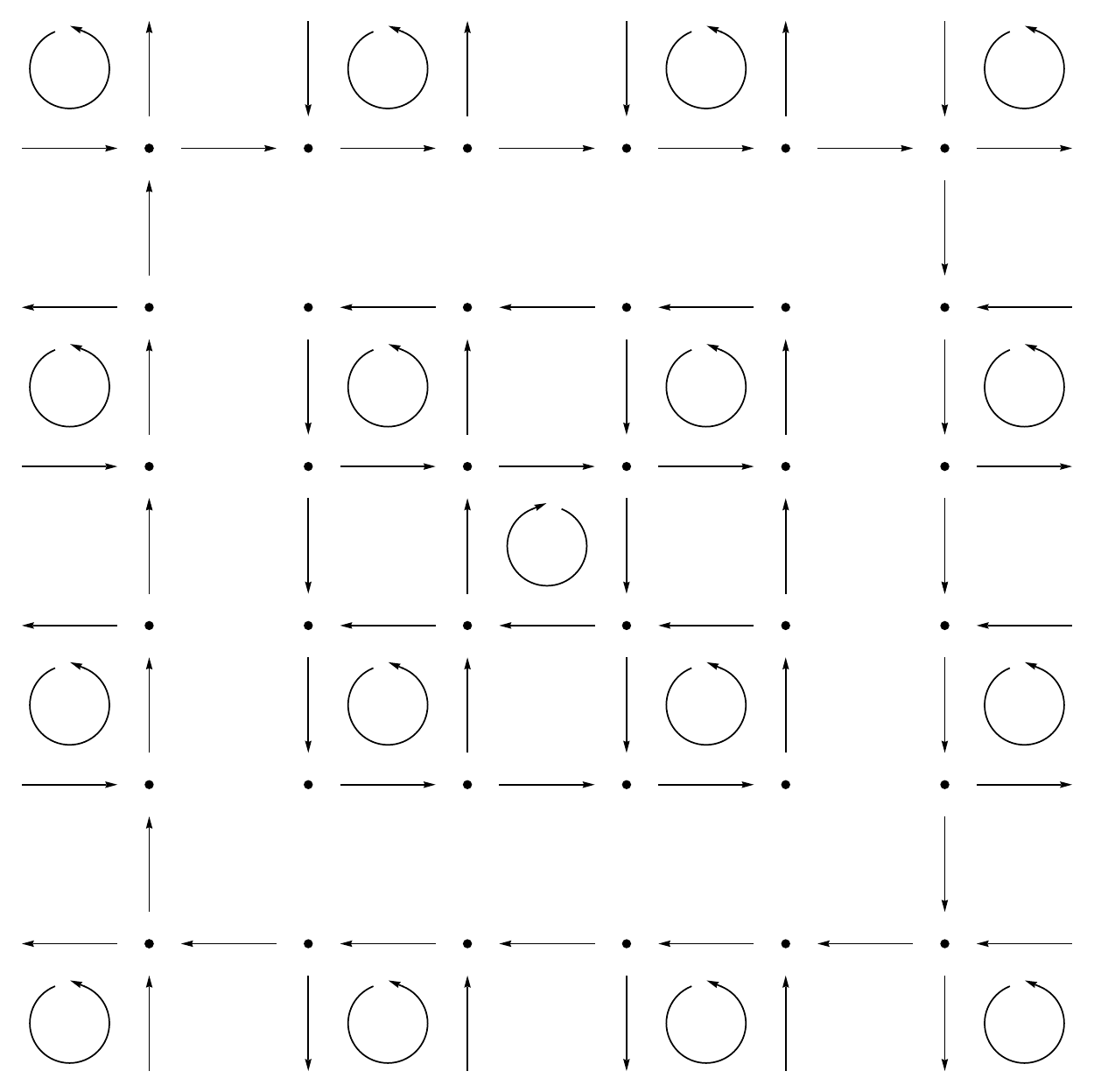}\\
\includegraphics[width=.5\textwidth]{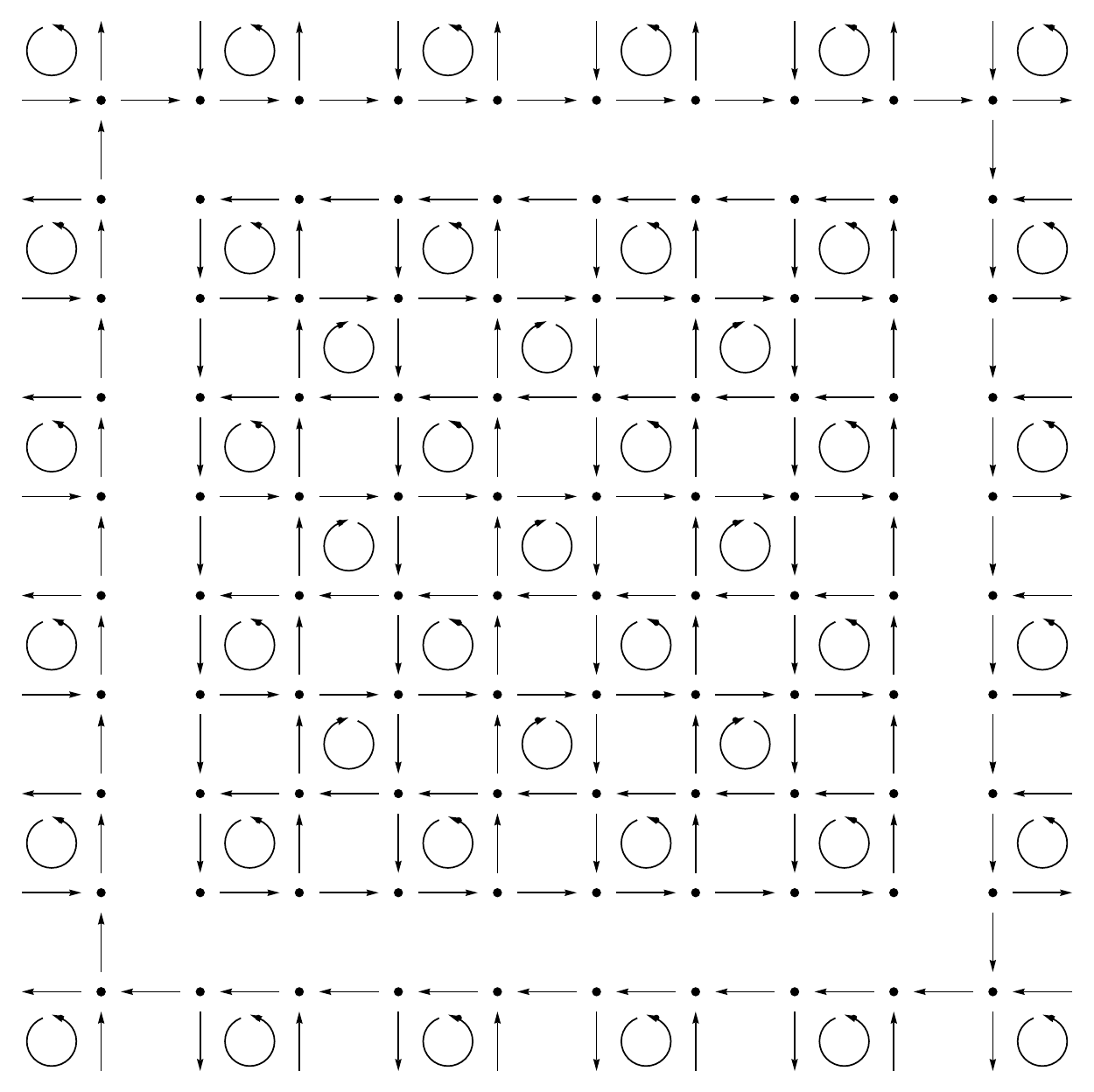}
}
 \caption{The action of $U^{L}$ for $L=(1,1)$ and $L=(2,2)$}
 \end{figure}

For $L\in\bZ^{2}$ and $\ii$ note the resolvent identities for $U^{(L)}$ and $U^{(L+\ii)}$ (omitting the dependences on the other variables)

\[\begin{split}
R&= R^{(L)}-R^{(L)}V^{(L)}R\\
&=R^{(L)}-R^{(L)}V^{(L)}\left(R^{(L+\ii)}-R V^{(L+\ii)}R^{(L+\ii)}\right).
\end{split}\]
Consult figure (\ref{fig:iota}).
We see that  for $\nu\in\stackrel{\circ}{\Lambda_{L+\ii}^{c}}$ it holds

\[\begin{split}&\langle e_{0},R e_{\nu}\rangle=\langle e_{0},R^{(L)}V^{(L)}R V^{(L+\ii)}R^{(L+\ii)}e_{\nu}\rangle\\
&=\sum_{
\substack{
\alpha\in\partial\Lambda_{L},\beta\in I_{\alpha}^{L}\\ 
\delta\in\partial\left({\Lambda}_{L+\ii}^{c}\right),\gamma\in O_{\delta}^{L+\ii}
}
}
\begin{split}\langle e_{0},R^{(L)}e_{\alpha}\rangle&\langle e_{\alpha},V^{(L)}e_{\beta}\rangle\langle e_{\beta},R e_{\gamma}\rangle\cdot \\&\cdot\langle e_{\gamma},V^{(L+\ii)}e_{\delta}\rangle\langle e_{\delta},R^{(L+\ii)}e_{\nu}\rangle\end{split}
\end{split}\]

with the definitions

$I_{\alpha}^{L}:=\lbrace\eta\in\bZ^{2}, \langle e_{\alpha},V^{(L)}e_{\eta}\rangle\neq0\rbrace,\quad
O_{\delta}^{L}:=\lbrace\eta\in\bZ^{2},\langle e_{\eta},V^{(L)}e_{\delta}\rangle\neq0\rbrace$.
Remark that the cardinality of each of these sets is not greater than two. From the boundedness of $V^{(L)}$ it follows the existence of a $c>0$ such that
\[\begin{split}\bE\left(\vert\langle e_{0},R e_{\nu}\rangle\vert^{s}\right)&\le \\& 
c\sum_{
\substack{
\alpha\in\partial\Lambda_{L},\beta\in I_{\alpha}^{L}\\ 
\delta\in\partial{\Lambda}_{L+\ii}^{c},\gamma\in O_{\delta}^{(L+\ii)}
}
}\bE\left(\vert\langle e_{0},R^{(L)}e_{\alpha}\rangle\vert^{s}
\vert\langle e_{\beta},Re_{\gamma}\rangle\vert^{s}
\vert\langle e_{\delta},R^{(L+\ii)}e_{\nu}\rangle\vert^{s}\right)\end{split}.\]

Now using the independence of the random variables $\vert\langle e_{0}, R^{(L)}e_{\alpha}\rangle\vert^{s}$ and $\vert\langle e_{\delta},, R^{L+\ii}e_{\nu}\rangle\vert^{s}$, equation (\ref{eq:momentestimate}), and the resampling argument as described in \cite{hjs}, p.435 ff. it follows for a $c>0$
\begin{equation}\label{eq:13.1}
\bE\left(\vert\langle e_{0},R e_{\nu}\rangle\vert^{s}\right)\le c\sum_{\alpha\in\partial\Lambda_{L}}\bE\left(\vert\langle e_{0},R^{(L)}e_{\alpha}\rangle\vert^{s}\right)\sum_{\delta\in\partial\Lambda_{L+\ii}^{c}}\bE\left(\vert\langle e_{\delta},R^{(L+\ii)}e_{\nu}\rangle\vert^{s}\right).
\end{equation}
Next use the resolvent identity
\[R^{(L+\ii)}=R+R^{(L+\ii)}V^{(L+\ii)}R\]
to estimate for $\delta\in\partial\Lambda_{L+\ii}^{c}$  and a $c>0$

\[\begin{split}\bE\left(\vert\langle e_{\delta}, R^{(L+\ii)} e_{\nu}\rangle\vert^{s}\right)&\le\bE\left(\vert\langle e_{\delta},R e_{\nu}\rangle\vert^{s}\right)+\\
&c\sum_{\alpha\in\partial\Lambda_{L+\ii}^{c},\beta\in I_{\alpha}^{L+\ii}}\bE\left(\vert\langle e_{\delta},R^{L+\ii}e_{\alpha}\rangle\vert^{s}\vert\langle e_{\beta},R e_{\nu}\rangle\vert^{s}\right).
\end{split}\]

Now employ a second tricky resampling argument as described in \cite{hjs}, p. 439f, equations (\ref{eq:momentestimate}) and (\ref{eq:13.1}) to estimate for a $c>0$

\begin{equation}
\bE\left(\vert\langle e_{0}, R  e_{\nu}\rangle\vert^{s}\right)\le c \vert L\vert \sum_{\alpha\in\partial\Lambda_{L}}\bE\left(\vert\langle e_{0}, R^{(L)}e_{\alpha}\rangle\vert^{s}\right) \sum_{\beta\in N\left(\partial\left(\Lambda_{L+\ii}^{c}\right)\right)}\bE\left(\vert\langle e_{\beta},R e_{\nu}\rangle\vert^{s}\right)\label{eq:13.2}
\end{equation}

where $N\left(\partial\Lambda_{L+\ii}^{c}\right)$ is the nearest neighborhood of $\partial\Lambda_{L+\ii}^{c}$.

 $\# \left(\partial\Lambda_{L}\right)=\cO(\vert L\vert)$ and $\#N\left(\partial\Lambda_{L+\ii}^{c}\right)=\cO(\vert L\vert)$ thus by proposition \ref{propo:resolventestimate} for any $a\ge0$, $\varphi$ small enough,  $\vert L\vert$ large enough there exists a $c>0$ such that
\[\bE\left(\vert\langle e_{0}, Re_{\nu}\rangle\vert^{s}\right)\le c\vert L\vert^{3-a}\max_{\beta\in N\left(\partial\Lambda_{L+\ii}^{c}\right)}\bE\left(\vert\langle e_{\beta},R e_{\nu}\rangle\vert^{s}\right).\]
Now choose $b>3$ and $\ell_{0}$ such that for $\lfloor\vert L\vert\rfloor=\ell_{0}$
\[c\vert L \vert^{3-b}=q<1\]
and the assertion is proved with the abuse of notation $L``=''L+{\ii}$.\qquad\ep

\begin{theorem}\label{thm:exponential}
For $s\in(0,1)$ there exist positive numbers $\varphi_{0}, c, g$ such that for all $\vert\varphi\vert\le\varphi_{0}$ and $\mu, \nu\in\bZ^{2}$, $z\in\bC\setminus\bT$
\begin{equation}
\bE\left(\vert\langle e_{\mu},R(\varphi,z) e_{\nu}\rangle\vert^{s}\right)\le c e^{-g\vert\mu-\nu\vert}
\end{equation}
\end{theorem}

{\bf Proof.} Refer to proposition \ref{propo:exponential}. By (\ref{eq:momentestimate}) it is sufficient to prove the claim for $\mu, \nu$ such that $\vert\mu-\nu\vert>\ell_{0}+2$.

\[\bE\left(\left\vert\langle e_{\mu}, R(\varphi,z) e_{\nu}\rangle\right\vert^{s}\right)\le q \max_{\beta\in N\left(\partial\left((\Lambda_{L}+\lbrack\mu\rbrack)^{c}\right)\right)}\bE\left(\left\vert\langle e_{\beta}, R(\varphi,z) e_{\nu}\rangle\right\vert^{s}\right)\]

for $q<1$ and $\vert L\vert$ big enough use the estimate again, replacing $\mu$ by $\beta$, then, iteratively, at least $\lfloor\frac{\vert\mu-\nu\vert}{\ell_{0}+2}\rfloor=:n$ times and use (\ref{eq:momentestimate}) for the last step to conclude that for a $c>0$
\[\bE\left(\vert\langle e_{\mu},R(\varphi,z) e_{\nu}\rangle\vert^{s}\right)\le c q^{n}\]
which proves the assertion by defining $g:=\frac{\vert\log q\vert}{\ell_{0}+2}$. \qquad\ep

The proof of theorem \ref{thm:main} consists now remarking that the estimate in theorem \ref{thm:exponential} implies exponential localization, theorem \ref{thm:main}(3) which in turn implies \ref{thm:main}(2) and \ref{thm:main}(1). These fact were proven in \cite{hjs}, theorem 3.2, propositions 3.1 and 3.2.

\section*{Acknowledgments}
We  acknowledge gratefully support from the
grants Fondecyt Grant 1080675; MATH-AmSud, 09MATH05; Scientific Nucleus
Milenio ICM P07-027-F, ECOS-CONICYT C10E01.

\end{document}